\documentclass[aps,prl,twocolumn,showpacs]{revtex4}

\usepackage{graphicx}
\usepackage{latexsym} 
\usepackage{amsmath,amscd}
\usepackage{times}
\usepackage{epstopdf}

\begin{document}

\title{\bf\noindent Universal Record Statistics of Random Walks and L\'evy Flights}
\author{Satya N. Majumdar$^1$ and Robert M. Ziff$^2$}
\affiliation{
$^1$ Laboratoire de Physique Th\'eorique et 
Mod\`eles Statistiques (UMR 8626 du CNRS), 
Universit\'e Paris-Sud, B\^at.\ 100, 91405 Orsay Cedex, France\\
$^2$  Michigan Center for
Theoretical Physics and Department of Chemical
Engineering, University of Michigan, Ann Arbor, MI USA 48109-2136}
\pacs{02.50.-r, 02.50.Sk, 02.10.Yn, 24.60.-k, 21.10.Ft}

\begin{abstract}
It is shown that statistics of records 
for time series generated by random walks are independent of
the details of the jump distribution, as long as the latter is continuous and symmetric.
In $N$ steps, the mean of the record distribution
grows as the $\sqrt{4N/\pi}$ while the standard
deviation grows as $\sqrt{(2-4/\pi) N} $, so the distribution
is non-self-averaging.
The mean shortest and longest 
duration records grow as  $\sqrt{N/\pi}$ and $0.626508... N$, 
respectively.  The case of
a discrete random walker is also studied, and  similar asymptotic behavior is found. 
\end{abstract}
\maketitle
The study of record statistics is an integral part of diverse fields including
meteorology~\cite{climate,RP}, hydrology~\cite{Matalas}, economics~\cite{Barlevy}, 
sports~\cite{GTS,Glick,BRV} and entertainment 
industries among 
others. 
In popular media such as television or newspapers, one always hears and reads about record 
breaking events. It is no
wonder that {\em Guinness Book of Records} has been a world's best-seller since 1955. 
In physics, records are relevant in the theory of domain-wall dynamics \cite{ABBM}, for example.
Consider any discrete time series  $\{x_0,x_1,x_2,\ldots,x_N\}$ of $N$ entries that may 
represent, e.g., 
the daily temperatures in a city or the stock prices of a company or the budgets
of Hollywood films. A {\em record} happens at step $i$ if the $i$-th entry $x_i$
is bigger than all previous entries $x_0$, $x_1$, $\ldots$, $x_{i-1}$.
Statisical questions that naturally arise are: (a) how many records occur in time 
$N$? (b) How long does a record survive? (c) what is the age of the longest surviving record?
etc. Understanding these aspects of record statistics is particularly important in the 
context of current issues of climatology such as global warming.

The mathematical theory of records has been studied for over 50 
years~\cite{Chandler,Nevzorov,ABN,SZ} and the questions posed in the previous
paragraph are well understood in 
the case when the random variables $x_i$'s are independent and identically
distributed (iid). 
Recently, there has been a resurgence of interest
in the record theory due to its multiple applications in diverse
complex systems such as spin glasses~\cite{SG}, adaptive processes~\cite{Orr} and  
evolutionary models of biological
populations~\cite{Krug1,Evol}. The results in the record theory of iid variables have
been rather useful in these different contexts.
Recently, Krug has studied the record statistics when the entries have non-identical
distributions but still retain their independence~\cite{Krug2}. 
However, in most realistic situations the entries of the time series are
{\em correlated}.
Surprisingly, very little is known about the statistics of records for a correlated time 
series. In this Letter we take a step towards this goal.

Of correlated  
time series $\{x_0, x_1, x_2,\ldots, x_N\}$, perhaps
the simplest and yet the most common  with a variety of 
applications~\cite{Feller}, 
is the one where
$x_i$ represents the position of a random walker at discrete time $i$. The 
walker
starts at $x_0$ at time $0$ and at each discrete step evolves via $x_i= x_{i-1} +\eta_i$
where the noise $\eta_i$ represents the jump length at step $i$. The jump lengths $\eta_i$'s 
are iid variables each drawn from a symmetric distribution $\phi(\eta)$. 
This also includes L\'evy flights where $\phi(\eta)\sim |\eta|^{-1-\mu}$ is
power-law distributed for large $|\eta|$ with exponent $0<\mu\le 2$ and thus has a
divergent second moment.
Even though the jump lengths are uncorrelated, the entries
$x_i$'s are clearly correlated.
This time series corresponding to a discrete-time Brownian motion appears naturally in
many different contexts. For example, in the context of  
queuing theory~\cite{queue}, $x_i$ represents the length of a single queue at time $i$.
In the context of the evolution of stock prices $x_i$ represents the logarithm
of the price of a stock at time $i$~\cite{finance}.  
In this Letter, we compute exactly the statistics of the number and the ages 
of records in this correlated sequence and show that the record statistics is universal,
i.e., independent of the noise distribution $\phi(\eta)$ as long as $\phi(\eta)$
is symmetric and continuous.  

It is useful to summarize our main results.  The record statistics are independent
of the starting position $x_0$ and hence without any loss of generality
we will set $x_0=0$ and also count the initial entry $x_0=0$ as the first record.
We show 
that the probability $P(M,N)$ 
of $M$ records in $N$ steps ($M\le N+1$) 
is simply
\begin{equation}
P(M,N)= \binom{2N-M+1}{N}\, 2^{-2N+M-1}
\label{nore1}
\end{equation}
which is  universal for all $M$ and $N$. The moments are also naturally universal
and can be computed for all $N$. In particular, for large $N$, the mean and the
variance behave as
\begin{eqnarray}
\langle M\rangle &\sim&  \frac{2}{\sqrt{\pi}}\sqrt{N} \nonumber \\
\langle M^2\rangle-{\langle M\rangle}^2 &\sim & 2\left(1-\frac{2}{\pi}\right)\, N
\label{mom1}
\end{eqnarray}
while the skewness, defined as the third central moment divided by the variance raised
to the 3/2-power, goes to a constant value $4(4-\pi)(2 \pi - 4)^{-3/2}$.
We also show that the age statistics of the records is universal for all $N$.
Evidently, the mean age of a  typical record grows, for large 
$N$, 
as $\langle l\rangle \sim N/\langle M\rangle\sim  \sqrt{\pi 
N/4}\approx 0.8862\,\sqrt{N}$. We also
compute the extreme age statistics, i.e., ages of the records
that have respectively the shortest and the longest duration.
These extreme statistics are also universal.
While the mean longevity of the record with the shortest age grows, for large $N$, as 
$\langle 
l_{\rm min}\rangle \sim \sqrt{N/\pi}\approx 0.5642\, \sqrt{N}$,  
that of the longest age  
grows faster, $\langle l_{\rm max}\rangle \sim c\, N$ where
$c$ is a nontrivial universal constant 
\begin{equation}
c = 2\int_0^{\infty} dy\, \log \left[1+ \frac{1}{2\sqrt{\pi}}
\,\Gamma(-1/2,y)\right]= 0.626508\ldots
\label{cons1}
\end{equation}
where $\Gamma(-1/2,y)= \int_y^{\infty}dx\, x^{-3/2}\, e^{-x}$. The universality
of these results can be traced back to the Sparre Andersen theorem 
on the first-passage property of random walks.

Let us consider any realization of the random walk sequence $\{x_0=0, x_1, x_2,\ldots, 
x_N\}$  (see Fig.\ 1), where $x_i=x_{i-1}+\eta_i$ 
and $\eta_i$'s are iid variables each drawn from the distribution $\phi(\eta)$. Let $M$ 
be the 
number of records in this realization.
Let ${\vec l}=\{l_1,l_2,\ldots, l_M\}$ denote the time intervals between successive records.
Thus $l_i$ is the age of the $i$-th record, i.e., it denotes the time up to
which the $i$-th record survives. Note that the last record, i.e., the $M$-th record,
still stays a record at the $N$-th step since there are no more record breaking 
events after it. Our aim is to first calculate the joint probability distribution
$P\left(\vec l, M|N\right)$ of the ages $\vec l$ and the number $M$ of records, given 
the
length $N$ of the sequence. For this, we need two quantities as inputs.
First, let $q(l)$ denote the probability that a walk, starting initially at $x$,
stays above (or below) its starting position $x$ up to step $l$. 
Clearly $q(l)$ does not depend on the starting position $x$.
A nontrivial theorem due to Sparre  Andersen~\cite{SA} states that 
$q(l)=\binom{2l}{ l} 2^{-2l}$ 
is universal for all $l$, i.e., independent
of $\phi(\eta)$ as long as $\phi(\eta)$ is symmetric and continuous.
Its generating function is simply
\begin{equation}
{\tilde q}(z)= \sum_{l=0}^{\infty} q(l)\, z^l = \frac{1}{\sqrt{1-z}}.
\label{SA1}
\end{equation}   
Our second input is the first-passage probability $f(l)$ that the walker crosses its 
starting
point $x$ for the first time between steps $(i-1)$ and $i$. Evidently, $f(l)= 
q(l-1)-q(l)$ with $l\ge 1$ is also universal and its generating function is
\begin{equation}
{\tilde f}(z)= \sum_{l=1}^{\infty} f(l) z^l = 1-(1-z){\tilde q}(z)=1-\sqrt{1-z}.
\label{SA2}
\end{equation}
\begin{figure}
\includegraphics[width=.9\hsize]{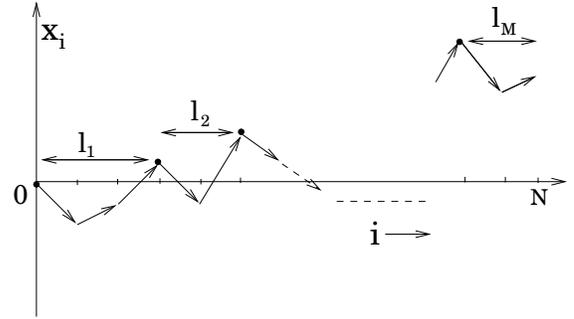}
\caption{A realization of the random-walk sequence $\{x_0=0,x_1,x_2,\ldots, x_N\}$ of 
$N$ 
steps with $M$ records.
Records are shown as black dots. $\{l_1,l_2,\ldots, l_M\}$ denotes the time intervals
between successive records.}
\label{figrw}
\end{figure}

Armed with these two ingredients $q(l)$ and $f(l)$, one can then write down 
explicitly the joint distribution of the ages $\vec l$ and the number $M$ of records 
\begin{equation}
P\left(\vec l, M|N\right)= f(l_1)\,f(l_2)\,\ldots f(l_{M-1})\,q(l_M)\, 
\delta_{ {\sum_{i=1}^M l_i,\, N}}
\label{joint1}
\end{equation}
where we have used the Markov property of random walks which dictates that
the successive intervals are statistically independent, subject to the
global sum rule that the total interval length is $N$ (see Fig.\ 1).
Note that since the $M$-th record
is the last one (i.e., no more records have  happened after it), 
the interval to its right has distribution $q(l)$ rather
than $f(l)$. One can check that $P\left(\vec l, M|N\right)$ is normalized to unity
when summed over $\vec l$ and $M$. Since $q(l)$ and $f(l)$ are universal
due to the Sparre Andersen theorem, it follows that $P\left(\vec l, M|N\right)$
and any of its marginals are also universal.

Let us first compute the probability of the number of records $M$, $P(M|N)=\sum_{\vec 
l} P\left(\vec l, M|N\right)$. To perform this sum, it is easier to consider its 
generating 
function. Multiplying Eq.\ (\ref{joint1}) by $z^N$ and summing over $\vec l$, one
gets
\begin{equation}
\sum_{N=M-1}^{\infty} P(M|N) z^N= [{\tilde f}(z)]^{M-1} 
{\tilde q}(z)=\frac{(1-\sqrt{1-z})^{M-1}}{\sqrt{1-z}}.
\label{nore2}
\end{equation}
By expanding in powers of $z$ and computing the coefficient of $z^N$, we get
our first result in Eq.\ (\ref{nore1}). One can also easily derive the
moments of $M$ from Eq.\ (\ref{nore2}). For example, for the first three moments we get
\begin{eqnarray}
\langle M\rangle &= &(2N+1)\,\binom{2N}{N}\, 2^{-2N} \nonumber \\
\langle M^2\rangle &=& 2N+ 2 - \langle M\rangle \nonumber \\
\langle M^3\rangle &=& -6 N - 6 + (7 + 4N) \langle M\rangle.   
\label{mom2}
\end{eqnarray}
The large-$N$ behavior in Eq.\ (\ref{mom1}) can then be easily derived from Eq.\
(\ref{mom2}) by using Stirling's approximation. In Fig.\ 2, we 
demonstrate this universality by computing from simulations $\langle M\rangle$
for three different distributions $\phi(\eta)$ (i) uniform in $[-1/2,1/2]$ (ii) Gaussian
with zero mean and unit variance 
and (iii) Cauchy or Lorentzian: $\phi(\eta)= {\pi}^{-1}/(1+\eta^2)$, which is 
an example of a L\'evy flight.
We then compare the data with the exact formula in Eq.\ (\ref{mom2}). 
The agreement is excellent and one cannot distinguish between the four
curves for any value of $N$.
\begin{figure}
\includegraphics[width=.9\hsize]{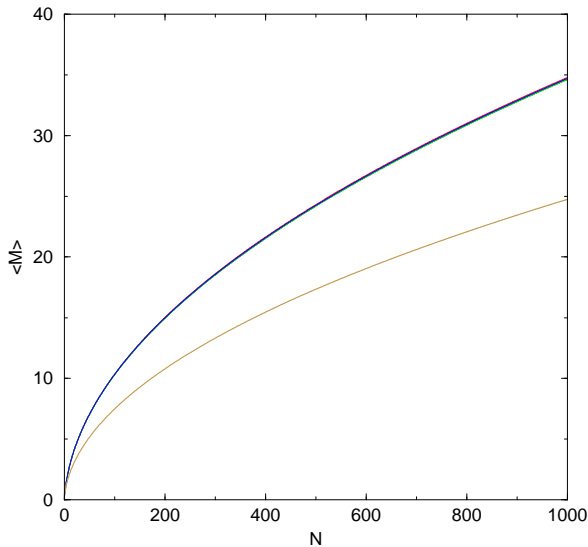}
\caption{(color online). The top curve actually contains four different curves denoting $\langle M\rangle$ 
vs $N$ for (i) uniform (ii) Gaussian (iii) Cauchy distributions for $\phi(\eta)$
and also (iv) the exact result in Eq.\ (\ref{mom2}). The four curves are
indistinguishable. The bottom curve shows $\langle M\rangle$ vs $N$ for
the lattice random walk with $\pm 1$ steps, i.e., when $\phi(\eta)= 
[\delta_{\eta,1}+\delta_{\eta,-1}]/2$, and agrees with the Eq.\ (\ref{discrete}).} 
\label{figavgrec}
\end{figure}

It is also interesting to compare this statistics of $M$ for the random-walk sequence
with that of the iid sequence where each entry $x_i$ is a random variable
drawn from some distribution $p(x)$. In the latter case, it is well
known~\cite{Nevzorov} that the distribution of the number of records $P(M|N)$ does not
depend on $p(x)$, and for large $N$, it approaches a Gaussian,
$P(M|N)\sim \exp[-(M-\log N)^2/{2\log N}]$, with
mean $\langle M\rangle = \log N$ and
the standard deviation $\sigma= \sqrt{\log N}$. Thus, fluctuations
of $M$ are small compared to the mean for large $N$. In contrast,
for the random-walk sequence, it follows from Eq.\ (\ref{mom1})
that both the mean and the standard deviation grow as $\sqrt{N}$
for large $N$ and thus the fluctuations are  large and comparable
to the mean. This suggests that in the random-walk case $P(M|N)$
has a scaling form for large $M$ and $N$, $P(M|N) \sim N^{-1/2}\,g(M N^{-1/2})$.
One can indeed prove this by analysing Eq.\  (\ref{nore2}) in the scaling limit
and finds $g(x)= e^{-x^2/4}/\sqrt{\pi}$.

While the  typical age of a record grows as $\langle l\rangle\sim N/\langle 
M\rangle\sim N^{1/2}$ for large $N$,
there are rare records whose
ages follow different statistics. For example, what is age distribution
of the longest lasting and the shortest lasting records? These extreme
statistics of ages can also be derived from the joint distribution
in Eq.\  (\ref{joint1}) and hence they are independent of $\phi(\eta)$.    

We first consider the longest lasting record with age $l_{\rm max}= {\rm 
max}(l_1,l_2,\ldots, l_M)$. It is easier to compute its cumulative distribution
$F(n|N)$, i.e., the probability that $l_{\rm max}\le n$ given $N$. 
Now, if $l_{\rm max}\le n$, it follows that  $l_i\le n$ for
$i=1,2,\ldots, M$. Thus, we need to sum up Eq.\  (\ref{joint1}) over all $l_i$'s
and $M$ such that $l_i\le n$ for each $i$. As usual it is easier to carry out this 
summation
by considering the generating function and we get
\begin{equation}
\sum_N F(n|N)\, z^N = \frac{\sum_{l=1}^n q(l) z^l}{1- \sum_{l=1}^n f(l) z^l}.
\label{fngf1}
\end{equation}
Extracting the distribution $F(n|N)$ from this general expression is somewhat
cumbersome and we do not present the details here~\cite{details}. However, one
can extract the asymptotic large-$N$ behavior of the average 
$\langle l_{\rm max}\rangle= \sum_{n=1}^{\infty} [1-F(n|N)]$ from
Eq.\  (\ref{fngf1}) using the explicit form of $q(l)$ and $f(l)$.
Skipping details~\cite{details}, we find that for large $N$, 
the mean age of the longest lasting record grows linearly with $N$,
$\langle l_{\rm max}\rangle 
\sim c N$ where $c= 0.626508\ldots$ is a universal constant given
in Eq.\  (\ref{cons1}). Thus, the age of the longest record ($\sim N$)
is much larger than the typical age ($\sim \sqrt{N}$) for large $N$.
Interesingly, exactly the same constant $c$ has appeared before in a different context 
\cite{PitmanYor97,Finch08}.

The statistics of the longest record for iid variables
follows
a similar asymptotic behavior $\langle l_{\rm max}\rangle \sim c_1 N$ but with the
prefactor~\cite{details} 
\begin{equation}
c_1= \int_0^{\infty} dx \exp\left[-x-\int_x^{\infty} dy \frac{e^{-y}}{y}\right]  = 0.624330\ldots
\end{equation}
which also describes the asymptotic linear growth of the longest cycle of
a random permutation and is known as the Golomb-Dickman
or Goncharov's constant (see \cite{Finch03}).
This result for iid variables also emerged recently in the context of a growing network 
model~\cite{GL}. Interestingly, the constant $c=0.626508..$  for random walks
is quite close to the Golomb-Dickman constant. 
It turns out that although the two problems (iid variables and random walks) have some 
common features
(at least qualitatively), the origin of universality is quite different in the 
two problems  \cite{details}.

For the record of the shortest duration $l_{\rm min}= {\rm min}(l_1,l_2,...l_M)$,
one find that the generating function of the 
cumulative distribution $G(n|N)$ denoting
the probability that $l_{\rm min}\ge n$ is given by
\begin{equation}
\sum_N G(n|N)\, z^N = \frac{\sum_{l=n}^{\infty} q(l) z^l}{1- \sum_{l=n}^\infty f(l) z^l}.
\label{gngf1}
\end{equation}
One can then extract, in a similar way, the asymptotic large-$N$ behavior 
of $\langle 
l_{\rm min}\rangle\sim \sqrt{N/\pi}$ \cite{details}. Thus, the mean age of the shortest lasting
record grows in a similar way as that of a typical record,
albeit with a smaller prefactor $1/\sqrt{\pi}=0.5642\ldots$ compared with
$\sqrt{\pi/4}=0.8862\ldots$, respectively.

\begin{figure}
\includegraphics[width=1.05\hsize]{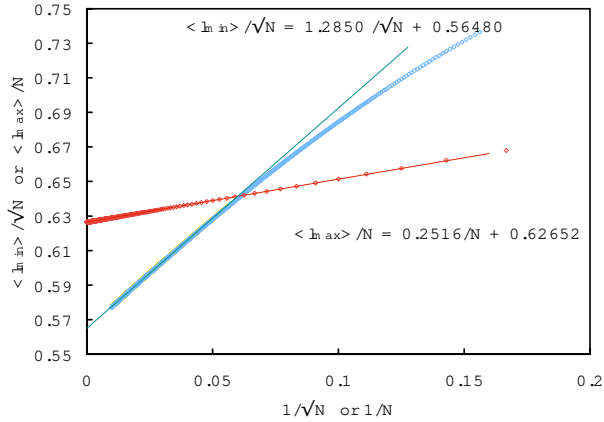}
\caption{(color online).  Plot of simulation results for 
$\langle  l_{\rm min}\rangle/\sqrt{N}$ vs.\ $1/\sqrt{N}$ (blue data falling on the steeper curve)
and  $\langle  l_{\rm max}\rangle/N$  vs.\ $1/N$ (red data falling on the less-steep curve), showing the asymptotic behavior of these two quantities.  Linear fits to the data for $500<N<10000$ yield
the straight lines, whose equations are displayed.  }
\label{figminmax}
\end{figure}

We have verified the results for  $\langle l_{\rm min}\rangle$ 
and  $\langle  l_{\rm max}\rangle$ numerically for the case of 
 jump distribution $\phi(\eta)$ uniform in $[-1/2,1/2]$, simulating 
 $10^9$ samples containing $10^4$ steps each.  We kept track
 of the largest and smallest interval between records (including the 
 final incomplete time interval)
 for each value of $N$, and calculated the average
 over all the runs.
 The results are shown in Fig.\ \ref{figminmax},
 where we plot $\langle  l_{\rm min}\rangle/\sqrt{N}$ and 
 $\langle  l_{\rm max}\rangle/N$, in the first case vs.\ $1/\sqrt{N}$, and in 
 the second case vs.\ $1/N$; making plots this way, we find that  the data falls
 on a nearly straight line as $N \to \infty$ in each case.   The intercepts, $0.56480$
 and $0.62652$, agree closely with the predictions, $\sqrt{1/\pi} = 0.564190\ldots$
 and $0.626508$, respectively.

We also considered the discrete (non-continuous) case where the walk
jumps by $\eta = \pm 1$ at each time step.  For this case
we find 
\begin{equation}
\sum_{N=0}^{\infty} \langle M \rangle z^N = \frac{\sqrt{1+z}+\sqrt{1-z}}{2(1-z)^{3/2}}
\end{equation}
which implies
\begin{equation}
\langle M \rangle = \frac{1}{2}\left[1+ \frac{ (-1)^{N+1}\Gamma(N-\frac{1}{2}) _2F_1(\frac{3}{2},-N;\frac{3}{2}-N;-1)}{2\sqrt{\pi}\Gamma(N+1)}\right]
\label{discrete}
\end{equation}
where $ _2F_1$ is the hypergeometric function, implying
 $\langle M \rangle = 1, 3/2, 7/4, 2, 35/16$, for $ N = 0, 1, 2, 3, 4$.  For large $N$,
$\langle M \rangle \sim \sqrt{2N/\pi}$, which is $1/\sqrt{2}$ of the expression for the mean in the continuous case.
We also find $\langle l_{\rm max}\rangle\sim c N$, and
$\langle l_{\rm min} \rangle \sim \sqrt{2N/\pi}$, which are respectively  equal to, and $\sqrt{2}$
times, the corresponding expressions for the continuous case.
These results were also verified in a simulation.

In conclusion, we have shown that the record statistics of a time series generated by a 
Markov process (random walk) are independent of the details of the walk distribution
when that distribution is continuous and symmetric.  Walks with a discrete jump 
distribution show similar asymptotic behavior but in general with different coefficients.
The results should be useful in analyzing a broad class of 
physical phenomena and are relevant for example to analyzing questions of 
climate change.   A possible future problem is the calculation of record statistics
for non-symmetric random jumps (with a drift) -- such as would be the case for a 
global warming trend.

Support of the National Science 
Foundation under Grant No. DMS-0553487 is gratefully acknowledged (RMZ).
Useful comments by Steven Finch are  highly appreciated.

\end{document}